\documentclass[prl,reprint,aps,showpacs]{revtex4-1}
\usepackage{dcolumn}
\usepackage[dvipdfm,colorlinks,urlcolor=blue,citecolor=blue,linkcolor=blue]{hyperref}
\usepackage[dvipdfm]{graphicx}
\usepackage{amsmath}
\usepackage{hypernat}
\begin{document}
\title{Characterization of hyperfine interaction between single electron and single nuclear spins in diamond assisted by quantum beat from the nuclear spin}

\author{J. H. Shim}
\altaffiliation{Present Address : Brain and Cognition Measurement Lab, Korea Research Institute of Standards and Science, Daejeon 350-340, Korea}
\author{B. Nowak}
\author{I. Niemeyer}
\author{J. Zhang}
\author{F. D. Brand\~{a}o}
\author{D. Suter}
\affiliation{
Fakult\"{a}t Physik, Technische Universit\"{a}t Dortmund, D-44221 Dortmund, Germany
}

\begin{abstract}
Precise characterization of a hyperfine interaction is a prerequisite for high fidelity manipulations of electron and nuclear spins belonging to a hybrid qubit register in diamond.
Here, we demonstrate a novel scheme for determining a hyperfine interaction, using single-quantum and zero-quantum Ramsey fringes, by applying it to the system of a Nitrogen Vacancy (NV) center and a $^{13}$C nuclear spin in the 1$^{\mathrm{st}}$ shell.
The zero-quantum Ramsey fringe, analogous to the quantum beat in a $\Lambda$-type level structure, particularly enhances the measurement precision for non-secular hyperfine terms. Precisions less than 0.5 MHz in the estimation of all the components in the hyperfine tensor were achieved. Furthermore, for the first time we experimentally determined the principal axes of the hyperfine interaction in the system. Beyond the 1$^{\mathrm{st}}$ shell, this method can be universally applied to other $^{13}$C nuclear spins interacting with the NV center.
\end{abstract}


\maketitle

The controllability of a nuclear spin in diamond, coupled to the electron spin of a single nitrogen vacancy (NV) center, is a crucial factor for its usefulness as a quantum resource.\cite{Dutt2007, Neumann2008, Jiang2009, Maurer2012} In most cases, it contributes to the decoherence, being a part of the nuclear spin bath.\cite{Balasubramanian2009, Lange2010, Lange2012} Under precise controls, however, nuclear spins turn into a quantum register that can be exploited as a quantum memory\cite{Dutt2007, Maurer2012, Fuchs2011, Shim2013} for a longer storage of quantum information, as auxiliary qubits\cite{Sar2012, George2013} for quantum algorithms, or possibly as a quantum magnetometer by forming a multipartite entangled state.\cite{Neumann2008, Jones2009}
Precise manipulations of electronic and nuclear spins require an accurate knowledge of hyperfine interactions;
otherwise, a microwave (mw) pulse designed to drive the electron spin can simultaneously change  the nuclear spin state.
A well-known example of this effect is the decay and revival of spin echoes of NV electron spins by the interaction with the $^{13}$C nuclear spin bath in diamond.\cite{Childress2006} Conversely, with a known hyperfine interaction, one may control indirectly nuclear spins via mw fields applied to the electron spin.
This approach provides significantly shorter gate durations\cite{Khaneja2007, Hodges2008}. Pulses designed by optimal control techniques could also exploit this effect to obtain a higher operation speed as well as a higher operation fidelity. Despite those potential impacts, only a few methods have been developed to determine the hyperfine interaction reliably.\cite{Childress2006} In this letter, we present a novel scheme aided by the quantum beat from the nuclear spin in the system. Our scheme provides a full characterization of a hyperfine interaction with uncertainties $<3\%$ for all  parameters.
We demonstrate the technique with a $^{13}$C nuclear spin in the 1$^{\mathrm{st}}$ coordination shell of a single NV center.

\begin{figure}[b]
\includegraphics[width=\columnwidth]{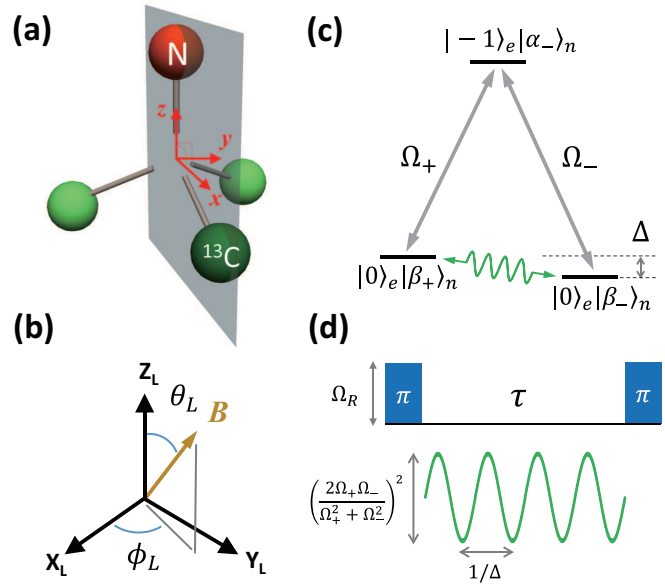}
\caption{(color online) \textbf{Symmetry of the system and zero quantum Ramsey sequence}
(a) The structure of the system, a NV center with a $^{13}$C nuclear spin in the 1$^{st}$ shell.
The plane containing the nitrogen, the vacancy, and the $^{13}$C is a symmetry plane.
The $x$, $y$, and $z$ axes in red define the NV frame, in which the hyperfine interaction is expressed.
(b) Orientation of external magnetic field in the Lab. frame.
(c) Energy level structure of the three lowest  states of the system.
The two ground states have $m_S=0$ and the excited state $m_S=-1$.
$\Omega_{\pm}$ represent transition amplitudes.
The green wave indicates zero quantum coherence between the two ground states.
(d) Zero quantum Ramsey sequence.
The oscillation amplitude depends on the ratio $\Omega_{+}/\Omega_{-}$, and its frequency is  $\Delta$,
the energy difference between the two ground states.}
\label{FIG1}
\end{figure}

To find the Hamiltonian of our system, we first consider the symmetry.
As illustrated in Fig.~\ref{FIG1} (a), the presence of the $^{13}$C nuclear spin reduces the point group symmetry
of the NV center from $C_{3v}$ to $C_s$, a single mirror plane.
This plane contains the $^{13}$C, the vacancy, and the nitrogen.
The symmetry-adapted axes consist of the $z$ axis along the axis of the NV
and the $x$-axis in the mirror plane, (red $x$, $y$, $z$ axes in Fig.~\ref{FIG1}(a)).
The presence of the mirror plane implies that all terms that are not invariant with respect to
the inversion of the $y$-coordinate in this frame must vanish.
This leaves us with the following form of the hyperfine Hamiltonian in the frequency unit ($h =1$):
\begin{equation}
\mathcal{H}_{\textrm{hf}} = A_{zz}S_z I_z + A_{xx}S_x I_x + A_{yy} S_y I_y +a (S_z I_x + S_x I_z).
\label{eqHam}
\end{equation}
Here, $S$ describes the electron spin,  $I$ the nuclear spin and $A_{\alpha \beta}$ the components of the
hyperfine tensor.
Since the hyperfine interaction is invariant with respect to the exchange of the spins \cite{Slichter}, we  set
$A_{zx} = A_{xz} = a$.
If the Hamiltonian is written in the principal axes system of the hyperfine tensor, $a$ vanishes;
here, we use the coordinate system of the NV center shown in  Fig.\,\ref{FIG1}. In contrast to earlier studies
\cite{Loubser1978, Bloch1985, Neumann2008, Gali2008, Felton2009}, which assumed  uniaxial symmetry $A_{xx} = A_{yy}$ in the principal axes system,
we do not impose any further symmetric constraints.

In principle, the parameters of Eq.(\ref{eqHam}) can be determined by measuring  spectra for different orientations of the magnetic field
and  fitting them numerically.\cite{Longdell2006}
The Ramsey fringe method generates spectra with the highest possible resolution and therefore optimal precision for the transition frequencies.
The precision $\delta A$ in the determination of the parameter $A$ is given by $\delta A = \delta \omega (\partial A/ \partial  \omega)$, in which $\delta \omega$ indicates the precision of the frequency $\omega$ and $c\{A\}=\partial \omega / \partial A$  the sensitivity with which the frequency depends on the parameter.
The frequency precision obtained from  a Ramsey fringe experiment with signal to noise ratio ($S/N$) and coherence time  $T_2^{*}$
is $ \delta \omega = 1 / (T_2^{\ast} \cdot S/N)$.
For  diamond crystals with natural abundance  $^{13}$C nuclei,  this precision is typically of the order of 0.2 MHz.
For the secular term $A_{zz}$, the parameter sensitivity is $c\{A_{zz}\} \approx 0.35$,
which results in a precision of $\delta A \approx 0.6$ MHz.
For the non-secular terms $A_{xx}$ and $A_{yy}$, however, the precisions are lower, since they contribute to the transition
frequency in 2$^{\mathrm{nd}}$ order, $\propto A^2 /D$.
As a result, the parameter sensitivities are  $c\{A_{xx} \} \approx 0.045$ and $c\{A_{yy}\} \approx 0.033$\cite{SI}.
The uncertainty of these parameters therefore becomes significantly larger, $\approx 5$ MHz.

In the following, we describe a procedure that reduces this uncertainty by about one order of magnitude.
It relies on the measurement of additional transition frequencies, which correspond (mostly) to
nuclear spin transitions, where the electron spin state does not change.
We refer to these transitions as  zero quantum transitions.
The advantage of this approach is that the coherence time of these transitions is an order of magnitude longer
than that of the single quantum transitions.\cite{SI}

We consider a system consisting of a single electron spin ($S=1$) and a $^{13}$C nuclear spin ($I=1/2$) as depicted in Fig.~\ref{FIG1}(a).
The Hamiltonian is
$\mathcal{H}_0=DS_z^2+\gamma_e \vec{B} \cdot  \vec{S} + \gamma_N \vec{B} \cdot  \vec{I} + \mathcal{H}_{\textrm{hf}}$.
$D$ is the zero-field splitting of the electron spin and
$\gamma_e$ and $\gamma_N$ are the gyromagnetic ratios of electron and nuclear spins.
The external magnetic field vector is $\vec{B}$ = $B(\sin{\theta} \cos{\phi}, \sin{\theta}\sin{\phi},\cos{\theta})$.
As shown in Fig.~\ref{FIG1}(c), the lowest three states of the system form a $\Lambda$-type level system.
The states $|\beta_{\pm}\rangle_n$ in Fig.~\ref{FIG1}(c) are superpositions of $|+1/2\rangle_n$ and $|-1/2\rangle_n$:
the effective quantization axis of the nuclear spin depends on the external magnetic field as well as on components of the hyperfine Hamiltonian.
For the excited state $|-1\rangle_e |\alpha_- \rangle_n$, in contrast, the quantization axis of the nuclear spin is dominated by
the hyperfine interaction ($>$ 100 MHz), which is much stronger than the nuclear Zeeman interaction ($<$ 0.5 MHz).
The quantization axis of the nuclear spins therefore is defined by the hyperfine Hamiltonian.
Its orientation is given by  $\theta'$, with $\tan \theta'=a/A_{zz}$ and the nuclear spin eigenstates are
$|\alpha_\pm \rangle = \pm \cos{\frac{\theta'}{2}}|\pm1/2\rangle+\sin{\frac{\theta'}{2}}|\mp1/2\rangle$.

Consequently, transition amplitudes are non-vanishing for the two transitions labeled $\Omega_\pm$ in Fig.~\ref{FIG1}(c).
The transition matrix elements of these transitions are proportional to the overlap $\langle \alpha_{-} | \beta_{\pm} \rangle$  of the nuclear spin states.
A microwave pulse  therefore couples to both transitions simultaneously and generates coherence
between the nuclear spin states, as marked by the green wave in Fig.~\ref{FIG1}(c).
A direct analogy to this situation is the excitation of  quantum beats in a $\Lambda$-type optical three level system.

The optimal excitation pulse for the nuclear spin coherence corresponds to an exchange of the `bright state' defined as
$|B\rangle$ $= \frac{1}{\sqrt{\Omega_+^2 + \Omega_-^2}} \left( \Omega_+ |0\rangle_e|\beta_+\rangle_n + \Omega_- |0\rangle_e|\beta_-\rangle_n\right)$
with the excited state.
\cite{SI, Note}
The resulting Ramsey-type signal can be written as
\begin{equation}
p(\tau) = \frac{\Omega_{+}^4 + \Omega_{-}^4}{(\Omega_{+}^2 + \Omega_{-}^2)^2} + 2 \left(\frac{\Omega_{+} \Omega_{-}} {\Omega_{+}^2+\Omega_{-}^2 }\right)^2 \cos{(\Delta \tau)},
\end{equation}
where $\Delta$ is the energy difference between  the ground states.
Using perturbation theory, we obtain an analytical expression for $\Delta$:\cite{SI}
\begin{equation}
\Delta = \frac{2 \gamma_e B \sin{\theta}}{D} \left( \sqrt{A_{xx}^2 + a^2} \cos^2{\phi} + A_{yy}\sin^2{\phi}\right).
\label{eqDelta}
\end{equation}
This expression shows that the zero quantum Ramsey signal can be exploited for the estimation of $\sqrt{A_{xx}^2 + a^2}$ and $A_{yy}$,
and that the orientation of the mirror plane in Fig.~\ref{FIG1}(a), which corresponds to $\phi = 0$,
can also be determined from the variation of  $\Delta(\phi)$.

\begin{figure}[t]
\includegraphics[width=\columnwidth]{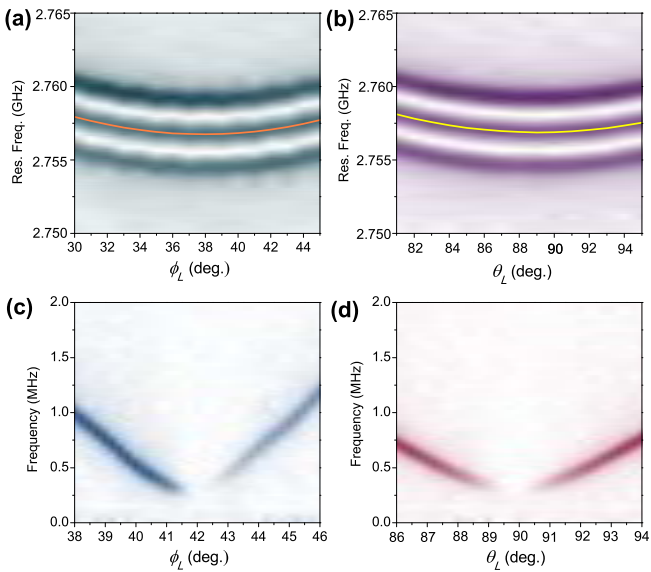}
\caption{(color online) \textbf{NV axis and single transition axis}
Single quantum Ramsey fringes were recorded while varying
the orientation of the external magnetic field and measuring a nearby NV center  with no first-shell $^{13}$C that has the same orientation.
The two angles $\theta_L$ and $\phi_L$ (see Fig. 1(b)) were independently scanned (a),(b)
The minima of the transition frequencies were reached at $\phi_L=37^{\circ}$ and $\theta_L=88.9^{\circ}$.
Zero quantum Ramsey fringes were measured on the system NV center to find the single transition axis. (c), (d)
The orientation of the single transition axis is $\phi_L=42^{\circ}$ and $\theta_L=89.8^{\circ}$.}
\label{FIG2}
\end{figure}

Microwave pulses always excite zero quantum coherence,
except when the magnetic field orientation is such that one of the two ground states becomes orthogonal
to the $|\alpha_-\rangle$ in the excited state.
We call this orientation single transition axis, since it results in a vanishing amplitude for one of the two transitions
in Fig.~\ref{FIG1}(c).

We start with the determination of the three terms, $D$, $B$, and $\sqrt{A_{zz}^2 + a^2}$,
which can be obtained from the single quantum Ramsey spectra.
This is possible for an arbitrary known orientation of the external magnetic field, but in practice, it is advantageous
to orient it along the single transition axis, because the Ramsey sequence then generates only single quantum coherences.
For the other orientations, a mixture of multiple coherences is produced,\cite{SI}, resulting in lower $S/N$.
Next, we determined the terms  $\sqrt{A_{xx}^2 + a^2}$ and $A_{yy}$ by performing a rotation of the magnetic field around the azimuthal angle $\phi$
in the NV frame and measuring the zero quantum Ramsey fringes.
The last step was the determination of $a$, which can be obtained from the orientation of the single transition axis
with respect to the $z$ axis in the NV frame.

The first experimental step is to find the orientation of the $z$-axis of the NV center and the single transition axis.
For this purpose, we rotated the magnetic field, keeping the field strength at the NV center constant, but varying the
orientation in the laboratory frame with a home-built 2D rotation stage\cite{SI}.
We describe the magnetic field orientation in the laboratory frame by the polar angles  $\theta_L$ and $\phi_L$.
The field-orientation maximizing the splitting between the energies of the
$m_S=\pm1$ states defines the $z$ axis.
We used a nearby NV center possessing the same crystallographic orientation and located within 5 $\mu$m from the system NV center,
and recorded single-quantum Ramsey fringes ($T_2^*$ $\approx$ 1 $\mu$s) as a function of the two angles.
Figure \ref{FIG2} (a) and (b) show the variation of the  frequency of the $m_S=0$ to $m_S=-1$ transition.
The three peaks at each angle arise from the hyperfine splitting with the $^{14}$N nuclear spin.
We took only the central transition to fit the angular variation, and the solid lines show the results.
The minimum of the fitted lines indicates the $\theta_L$ and the $\phi_L$ of the NV axis.
The single transition axis can be found in the same way from the zero quantum Ramsey fringes ($T_2^*$ $\approx$ 20 $\mu$s).
Fig.~\ref{FIG2}. (c), (d), show the variation of frequency and amplitude of the signals as a function of the two angles.
The results clearly shows that the zero quantum Ramsey fringes vanish at a certain orientation \cite{SI}.
The relative angle between the single transition axis and the NV axis is  5.1$^{\circ}$.

\begin{table}[b]
\caption{\label{table} All the parameters of the hyperfine interaction between a NV electron spin and a $^{13}$C nuclear spin in the 1st shell are shown.
Other parameters not shown are zero.
The principal values of the hyperfine interaction, $\overline{A}_{xx}$, $\overline{A}_{yy}$, and $\overline{A}_{zz}$, are presented.
$\theta_P$ is the angle between the two $z$ axes of the NV frame and  the principal axes system.}
\begin{ruledtabular}
\begin{tabular}{c|c||c|c}
$A_{xx}$ & 166.9($\pm$0.2)MHz & $\overline{A}_{xx}$ & 30.3($\pm$0.1)MHz\\
$A_{yy}$ & 122.9($\pm$0.2)MHz & $\overline{A}_{yy}$ & 122.9($\pm0.2$)MHz\\
$A_{zz}$ & 90.0($\pm$0.3)MHz & $\overline{A}_{zz}$ & 226.6($\pm$0.5)MHz \\
$A_{zx}$=$A_{xz}$ ($\equiv a$)& -90.3($\pm$0.3) MHz & $\theta_P$ & 56.5$(\pm$0.2)$^{\circ}$ or  123.5$^{\circ}$\\
\end{tabular}
\end{ruledtabular}
\end{table}

\begin{figure}[t]
\includegraphics[width=\columnwidth]{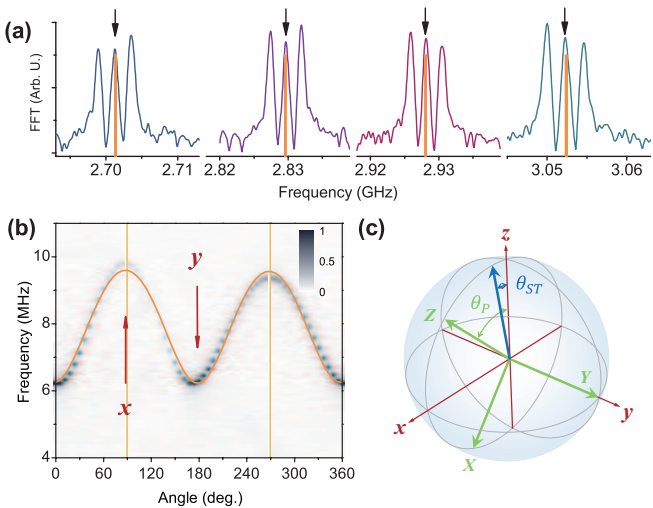}
\caption{(color online) \textbf{Single quantum spectrum and zero quantum spectra}
(a) Single quantum  spectrum for the magnetic field oriented along the single transition axis.
The three peaks of each transition are due to the hyperfine splitting of $^{14}$N nuclear spin.
The numerical analysis used the four transition frequencies marked by arrows and vertical bars.
(b) The zero quantum spectrum was measured as a function of the azimuthal angle ($\phi$) for ($\theta=40^{\circ}$).
The origin  $\phi=0$ is arbitrary and does not correspond to the $x$ axis.
The solid orange line represents the numerical fit. The yellow bars indicate the mirror plane in Fig.~\ref{FIG1}(a) obtained from the orientation of the single transition axis. (c) Orientations of the single transition axis (blue) and the principal axes (green) of the hyperfine tensor.
The $x$, $y$, $z$ axes in red present the NV frame, shown in Fig. 1(a).}
\label{FIG3}
\end{figure}

The single quantum spectra were obtained under the single transition axis using the Ramsey fringe sequence with $\pi/2$ pulses.
It has the main four single quantum transitions and each transition shows hyperfine splitting due to the $^{14}$N nuclear spin.
Since the hyperfine interaction with the $^{14}$N is not relevant for the present study,
we used the central transition frequency, which corresponds to the $m_I=0$ state of the $^{14}$N.
They are marked by arrows in Fig.~\ref{FIG3} (a).
Finally, an azimuthal rotation of the external magnetic field with respect to the orientation of the system NV center was performed,
and figure \ref{FIG3} shows the result for $\theta=40^{\circ}$.
Here, $\phi=0$ does not correspond to the $x$ axis in the NV frame but to an axis given by the configuration of the magnet rotation stage.
From the angular variation of the results, we can find the $x$ and $y$ axes in the NV frame, indicated by the red arrows. By definition, the $x$ axis lies in the mirror plane whose orientation is marked by the yellow vertical lines in Fig.~\ref{FIG3} (b).
The orientation of the symmetry plane can be determined independently from the orientation of the single transition axis, which must also lie in the symmetry plane.

Table 1 shows the hyperfine parameters determined from these experimental data.
The orange solid line in Fig.~\ref{FIG3}(b) is the best fit curve obtained from the numerical analysis,
and the orange bars in Fig.~\ref{FIG3}(a) represent the transition frequencies obtained.
The deviation between experimental and fitting results in Fig.~\ref{FIG3}(b) is attributed to the experimental imperfections in the rotation of the magnetic field,
which result in in a variation of the magnetic field strength.
We performed the same rotation on nearby NVs, and its deviation from g was similar to that of the system.\cite{SI}

From the hyperfine tensor estimated in the NV frame, one can find the principal axes in which the hyperfine tensor becomes diagonal.\cite{Slichter}
The principal values, $\overline{A}_{xx}$, $\overline{A}_{xx}$, and $\overline{A}_{zz}$, are summarized in  table.~\ref{table}, together with $\theta_P$,
the  angle between the $z$ axis in the NV frame and that of the principal axes system.
The ambiguity for the angle is a result of the fact that we can not decide if the positive $z$-axis points from the N to the V or vice versa.
These values are clearly inconsistent with earlier results based on a simpler version of the Hamiltonian.\cite{Loubser1978, Bloch1985, Neumann2008, Gali2008, Felton2009}

The experimental scheme presented can also be applied  to other $^{13}$C nuclear spins.
If the single quantum coherence decays on the time scale of $T_2^{\ast}$,
the resulting measurement precision is of the order $1/T_2^{\ast}$.
For distant $^{13}$C nuclear spins with weak hyperfine interactions, the number of decoherence sources,
such as nitrogen impurities or  $^{13}$C nuclear spins must be reduced\cite{Balasubramanian2009} until $T_2^{\ast} A > 1$.

The present study of the hyperfine characterization significantly improves the understanding of the hybrid qubit system consisting  of the NV electron spin
($S=1$) and a nuclear spin ($I=1/2$).
Our results show that the two level approach to describe mw and rf excitations is not applicable
because the excitation induces other transitions, which can lead to  unwanted operations.
In all  studies manipulating a single $^{13}$C nuclear spin in diamond,
the magnetic field should be orientated along the single transition axis,
as in our earlier work.\cite{Shim2013}
This becomes even more crucial in a system containing several nuclear spins.\cite{Neumann2008}
When a NV center has more than single $^{13}$C nuclear spin in the proximal sites,
it is not feasible to find a common single transition axis for all the $^{13}$C nuclear spins.
The secular approximation, therefore becomes invalid, and a complicated spin response can occur
even under single frequency excitations.

In conclusion, we demonstrated that the hyperfine interaction between a single NV electron and a $^{13}$C nuclear spin
can be characterized by a self-consistent method which  uses zero-quantum coherence from the $\Lambda$ structure
inherent in the system.
The long coherence time of the zero quantum coherence provides an enhanced precision for non-secular terms,
which is not feasible just by using single quantum coherences of the NV electron spin.
In contrast  to earlier attempts  for a better characterization of the hyperfine interaction,\cite{Jelezko2004, Felton2009, Smeltzer2011, Dreau2012}
our work yields the full hyperfine tensor and the secular approximation is not needed.
A precise information about the hyperfine interaction will allow us to adopt the optimal control method to implement a desired operation on the hybrid qubit system
with high fidelity.\cite{Palao2002}
This may include the creation of a multipartite entanglement among the $^{13}$C nuclear spins,
such as the NOON state for quantum metrology applications.
In addition, ZEFOZ (ZEro First Order Zeeman) points\cite{Longdell2006} can be used to obtain longer coherence times,
similar to atomic clock systems.
The anisotropy in the principal values of the hyperfine tensor could be exploited for a vector sensing of external magnetic field,
in contrast to the conventional projected field sensing using the NV center.\cite{Chen2013}

\acknowledgments
We gratefully acknowledge useful discussions with Dima Budker.
This work was supported by the Deutsche Forschungsgesellschaft through grant Su 192/27-1  (FOR 1482).

%


\onecolumngrid
\vspace{1.5cm}
\begin{center}
\textbf{\large Supplmentary Information}
\end{center}

\setcounter{figure}{0}
\setcounter{equation}{0}
\makeatletter
\renewcommand{\thefigure}{S\@arabic\c@figure}
\renewcommand{\theequation}{S\@arabic\c@equation}
\renewcommand{\bibnumfmt}[1]{[S#1]}
\renewcommand{\citenumfont}[1]{S#1}
\section{Experimental setup}
\subsection{Optical and microwave equipment}
Single NV-centers in diamond were optically addressed using our home-built confocal microscope. For the optical excitation and the detection of a single NV center, a diode-pumped solid state continuous wave (CW) laser with 532 nm wave length was used. For pulsed experiments, an acousto-optic modulator of 58 dB extinction ratio and of 50 ns rising time shaped laser pulses from the incident CW laser. An oil immersion microscope objective lens having 1.4 N.A. focused the exciation laser to single NV centers and collected fluorescences from single NV centers. The single photon detector (from PicoQuant) counted the number of photons from phonon-side band of NV centers' fluoresecence behind the long pass filter of 650 nm. The system NV center investigated in the present study showed a saturation fluorescence of 145 kcps under 1 mw excitation power.

A resonant mw frequency was generated by mixing an IF signal from Direct Digital Synthesizer (DDS) of 1 GS/s and a LO signal from a stable frequency source (APSIN 3000 from Anapico). An IQ mixer (IMOH-01-458 from Pulsar microwave) in combination with a quadrature hybrid (HYB01-500-06 from mitec) were adopted for the single side band mixing. The side band rejection of 42 dB was acheived after filtering the RF output of the IQ mixer with a band pass filter (2.7 - 3.1 GHz from minicircuit). The frequency and the phase of the IF was controlled via programming the DDS. The RF switch (ZASWA-2-50DR+ from minicircuit) modulated the mw amplitude for pulsed excitations. Through an amplier (ZHL-16W-43-S+ from minicircuit), the mw excitation pulse were guided to a Cu wire of 20 $\mu$m diameter attached on the surface of the diamond crystal. A digial pulse generator device (PulseBlaster ESR PRO from spin core) conducts all the timings by sending timing signals with a resolution of 2 ns (500 MHz clock). All the experiments and the analysis were performed on the software platform of Labview.

\subsection{Magnetic field rotation stage}
\begin{figure}[t]
\includegraphics[width=14cm]{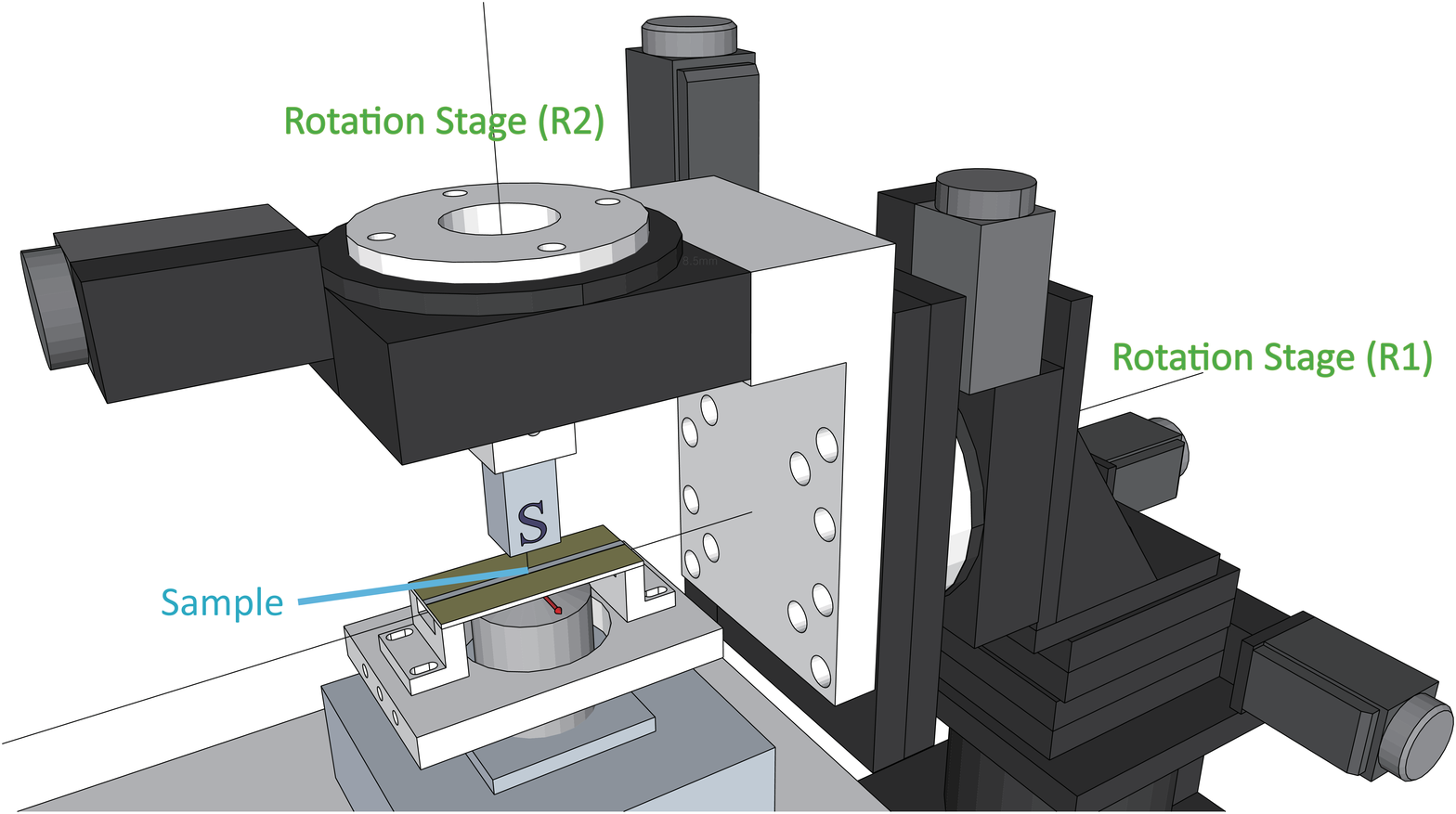}
\caption{(color online)\textbf{Rotation stage with a permanent magnet} The axes of the two rotation stage R1 and R2 intersect with each other at the position of the sample. The red arrow indicates the orientation of external magnetic field from the permanent magnet. The software Sketchup was used for designing.}
\label{FIGS1}
\end{figure}

For a full degree of freedom in the rotation of external magnetic field, we constructed a mechanical frame for a permanent magnet, which can independently vary the polar angles $\phi$ and $\theta$ in the Lab fame. It’s made of two motorized rotation stages (8MR151 from STANDA), which are backlash free, and three motorized linear stages (8MT167 from STANDA) and one manual vertical stage (SIG-122-0255 from Optosigma). The frame is designed to make the axes of the two rotation stages intersect with each other. This intersection point was adjusted to the position of the diamond crystal by using the three linear translation stages. In Fig.~\ref{FIGS1}, the rotation of the stage R1 corresponds to the $\phi$ variation and the stage R2 to the $\theta$ in the Lab frame. The red arrow at the position of the sample points the orientation of external magnetic field produced by the permanent magnet of a bar shape.

\section{Zero-quantum Ramsey fringe in a $\Lambda$-type spin level structure}
\subsection{Formation of bright and dark spin states}
\begin{figure}[t]
\includegraphics[width=12cm]{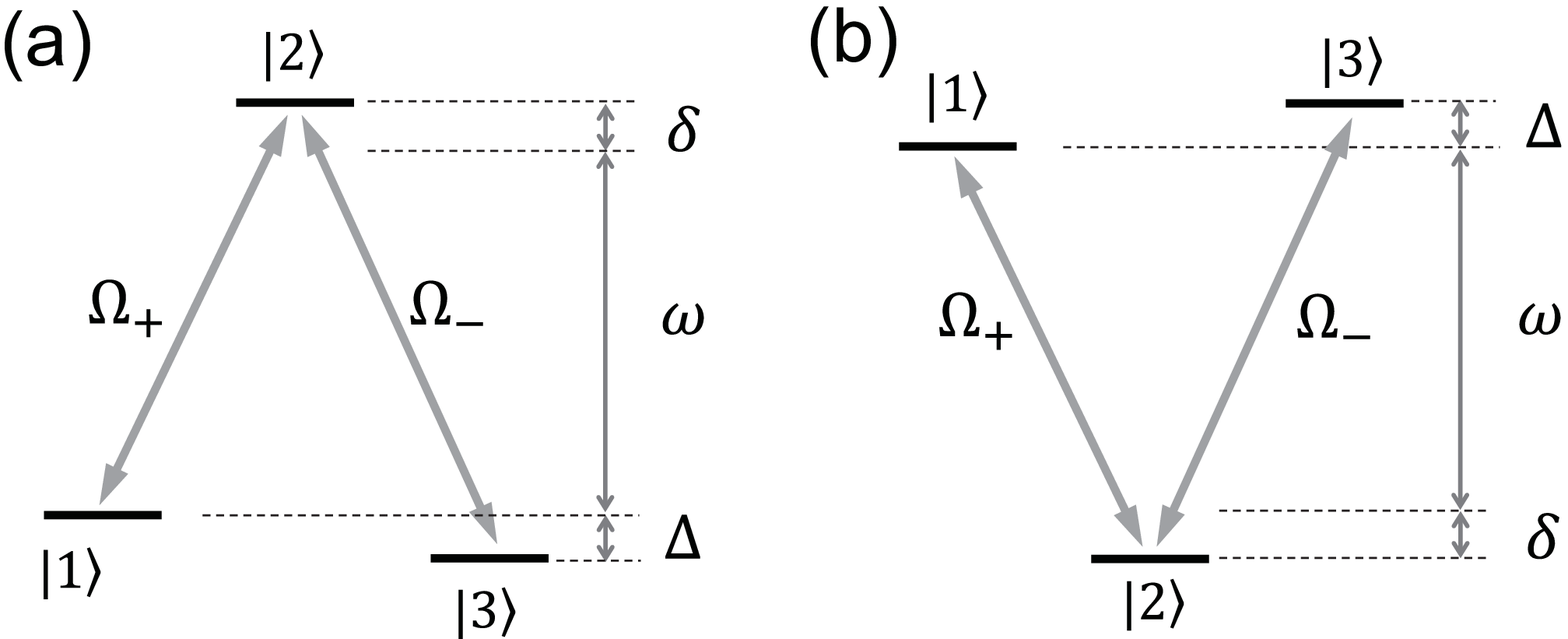}
\caption{(color online)\textbf{Models of a $\Lambda$ (a) and a V (b) level structures.} The actual system of a NV center with a $^{13}$C nuclear spin in the 1$^{\mathrm{st}}$ coordination has a $\Lambda$ level structure in it, but the consideration of the V level structure gives out an intuitive way of understanding the spin dynamics of the zero quantum Ramsey sequence. }
\label{FIGS2}
\end{figure}

In this section, we will mathematically present how the dark and the bright spin states can be formed under single-frequency mw excitation to the system of a $\Lambda$-type spin level structure. Figure~\ref{FIGS2}(a) illustrates the energy level structure of three states,$|1\rangle$, $|2\rangle$, and $|3\rangle$. We introduce Hamiltonian ($\mathcal{H}_0$) that includes the $\Lambda$-type spin level structure.  A mw field of a frequency $\omega$ is exposed to the system. Provided that the energy difference between the state $|1\rangle$ and $|3\rangle$, assigned as $\Delta$, is smaller than the mw strength, both transitions indicated as two arrows $\Omega_+$ and $\Omega_-$ will be induced simultaneously by the mw excitation. The total Hamiltonian $\mathcal{H}$ contains the system Hamiltonian $\mathcal{H}_0$ and the excitation field as
\begin{equation}
\mathcal{H} = \mathcal{H}_0 + \widetilde{\Omega}\cos{(\omega t)}.
\label{S1}
\end{equation}
The transition matrix $\widetilde{\Omega}$  can be represented in a matrix form as
\begin{equation}
\left(\begin{array}{ccc}
0 & \Omega_- & 0 \\
\Omega_+ & 0 & \Omega_- \\
0 & \Omega_- & 0 \\
\end{array}\right)
\begin{array}{c}
|1\rangle \\ |2\rangle \\ |3\rangle \\
\end{array}.
\end{equation}

By performing a unitary transformation $U$ to the system, which mixes the states $|1\rangle$ and $|3\rangle$, we can eliminate one of the transitions and find the corresponding bright $|B\rangle$ and dark $|D\rangle$ spin states.
\begin{equation}
U=\left(
\begin{array}{ccc}
\sin{\lambda} & 0 & -\cos{\lambda} \\
0 & 1 & 0 \\
\cos{\lambda} & 0 & \sin{\lambda} \\
\end{array}
\right)
\end{equation}
If the $\lambda$ satisfied the condition, $\tan{\lambda} = \frac{\Omega_-}{\Omega_+}$, the transformed transition matrix $U \widetilde{\Omega} U^{\dagger}$induces only the transition between the state $|2\rangle$ and the $|B\rangle$, and, thus, the $|D\rangle$ remains unreachable by the mw excitation.
\begin{equation}
U \widetilde{\Omega} U^{\dagger} = \left(
\begin{array}{ccc}
0 & 0 & 0 \\
0 & 0 & \sqrt{\Omega_+^2 + \Omega_-^2}\\
0 & \sqrt{\Omega_+^2 + \Omega_-^2} & 0 \\
\end{array}
\right),
\end{equation}
\begin{equation}
|B\rangle = \frac{1}{\sqrt{\Omega_+^2 + \Omega_-^2}} \left( \Omega_+ |1\rangle + \Omega_- |3\rangle\right),
|D\rangle = \frac{1}{\sqrt{\Omega_+^2 + \Omega_-^2}} \left( \Omega_- |1\rangle - \Omega_+ |3\rangle\right).
\end{equation}
In summary, in a $\Lambda$ level structure, unless $\Omega_+$ or $\Omega_-$ becomes zero, the bright and the dark states exist, and only the transition between the bright state $|B\rangle$  and the state $|2\rangle$ will be induced by mw excitation.

\subsection{Hamiltonian in the rotating frame}
The Hamiltonian in the Eq.(\ref{S1}) can be transformed into the rotating frame according to the frequency $\omega$ of the mw excitation. The $\mathcal{H}_0$ can be represented in a matrix form as
\begin{equation}
\mathcal{H}_0 = \left(
\begin{array}{ccc}
-(\omega + \delta) & 0 & 0 \\
0 & 0 & 0\\
0 & 0 & - (\omega + \delta + \Delta)\\
\end{array}
\right),
\end{equation}
where the $\delta$ is the detuning of the mw frequency with respect to the transition $\Omega_+$. By applying the rotating wave approximation, the Hamiltonian in the rotating frame ($\mathcal{H}_0^R$) can be simplified as,
\begin{equation}
\mathcal{H}_0^R = \left(
\begin{array}{ccc}
0 & 0  & 0\\
0 & +\delta & 0 \\
0 & 0 & - \Delta \\
\end{array}
\right).
\end{equation}

\subsection{Frequency and amplitude of Zero-quantum Ramsey}
\subsubsection{V - type level structure}
If the system has a $V$ level structure instead of the actual $\Lambda$ in the system, the mathematical description of zero quantum Ramsey becomes easier to understand. Hence, we start with the Zero quantum Ramsey from a $V$ level structure.(Figure~\ref{FIGS2} (b)) We assume that, prior to the zero quantum Ramsey sequence, the system state is initialized to the state $|2\rangle$ by a laser excitation and the population ($p$) of the state $|2\rangle$ will be measured from the fluorescence of the NV center.
After the first $\pi$ pulse of the sequence, the bright spin state $|B\rangle$ is formed, and then it will undergo a free evolution during the time $\tau$. The free evolution of the state $|B\rangle$ can be calculated by multiplying $e^{-i\mathcal{H}_0^R \tau}$ to the $|B\rangle$, and the resulting state is $|B(\tau)\rangle = \cos{\lambda}|1\rangle + e^{i\Delta \tau}\sin{\lambda}|3\rangle$. This information of in the relative phase will be transferred to the population of the $|2\rangle$ by the second $\pi$ pulse. The measured signal can be expressed as below
\begin{equation}
p(\tau) = |\langle B(\tau) | B\rangle|^2 = \frac{\Omega_+^4 + \Omega_-^4}{(\Omega_+^2+\Omega_2^2)^2} +
2\left(\frac{\Omega_+\Omega_-}{\Omega_+^2+\Omega_-^2}\right)^2 \cos{(\Delta \tau)}
\label{EqS8}
\end{equation}
The Eq. (\ref{EqS8}) is the mathematical form of the zero-quantum Ramsey from the $V$ level structure shown in Fig.\ref{FIGS2} (b). We can notice that the energy difference $\Delta$ can be measured from the oscillation of Zero-quantum Ramsey.

\subsubsection{$\Lambda$-type level structure}
Here, we consider the actual situation that the system owns the $\Lambda$ level structure shown in Fig.~\ref{FIGS2}(a). The state $|1\rangle$ and $|3\rangle$ correspond to the states $|0\rangle_e |\beta_+\rangle_n$ and $|0\rangle_e |\beta_-\rangle_n$  in Fig.1(c), respectively. The NV center is initialized into $m_S=0$ state prior to the zero-quantum Ramsey sequence. Since the population of $^{13}$C nuclear spin is fully mixed among the two states $|\beta_{\pm}\rangle$, in this model the states $|1\rangle$ and $|3\rangle$ are also equally populated initially. The population of a state $|i\rangle$ at time $t$ will be written as $p_i (t)$. The $p_{i|j} (t)$ indicates the conditional population of the state $|i\rangle$ if the initial state of the system is $|j\rangle$. At $t=0$, $p_1 (0)=p_3 (0)=1/2$.

The optical readout of NV center tells about the population in the $m_S=0$ states, $p(\tau)$. Therefore, the quantity one measures after the zero-quantum Ramsey sequence is $p_1 (t)+p_3 (t)$, which is equal to $1-p_2 (t)$. Since $p_2 (t)= p_{2|1} (t) p_1 (0)+p_{2|3} (t) p_3 (0)$, we will derive the expressions for $p_{2|1} (t)$ and $p_{2|3} (t)$ separately. First, the calculation steps for $p_{2|1} (t)$ proceed as below, in which $U_{\pi}$ represents the operation of $\pi$ pulses in the sequence, i.e., the population inversion between $|2\rangle$ and $|B\rangle$.

\begin{figure}[t]
\includegraphics[width=\textwidth]{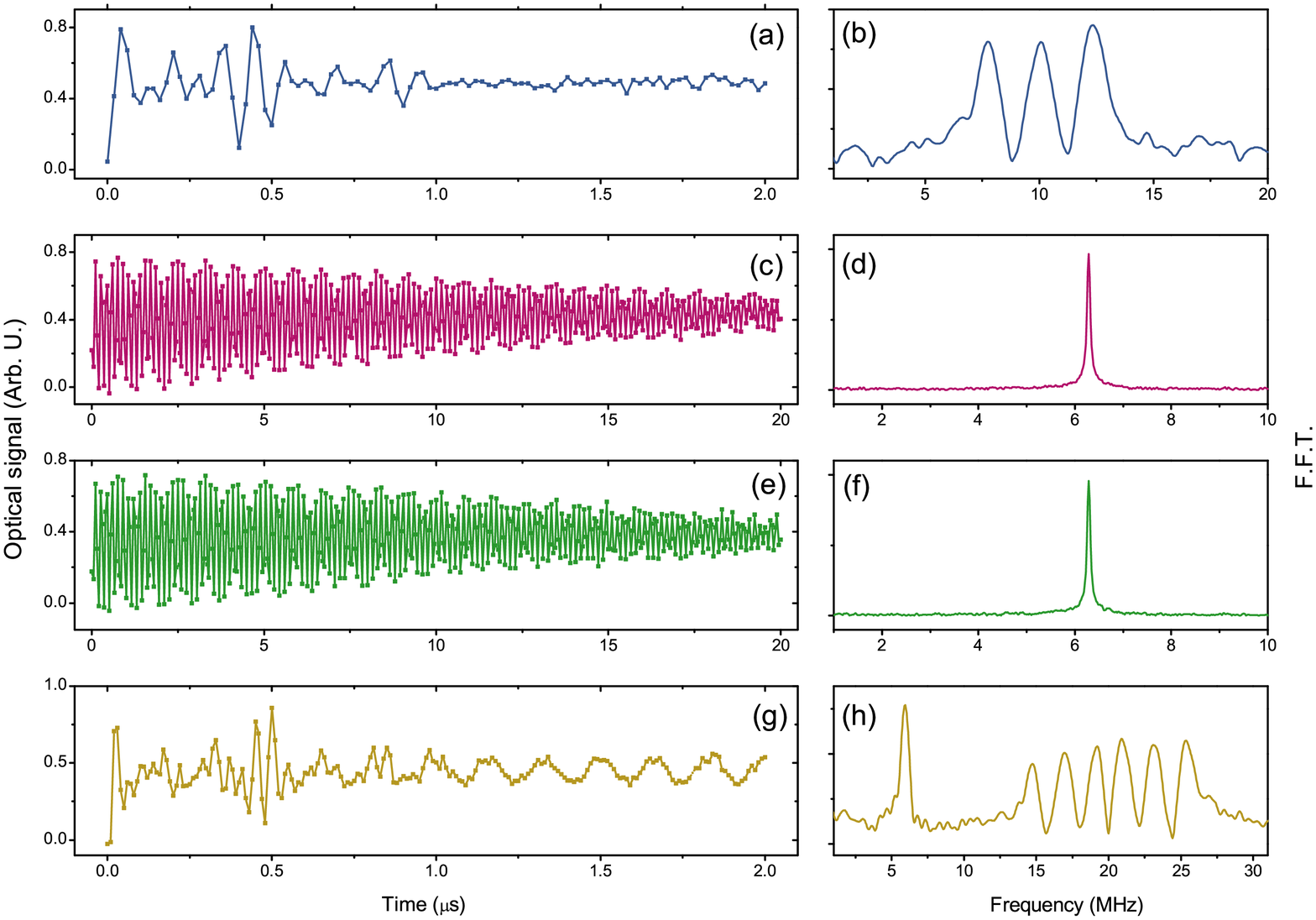}
\caption{(color online)\textbf{Comparison of single and zero quantum coherences} (a) The single quantum coherence measured under single transition axis. The coherence time is less than 1 µs. (b) The FFT of the curve in (a). The triplet stems from the hyperfine interaction between $^{14}$N nuclear spin ($I=1$) in the NV center. (c) The zero quantum coherence with detuning ($\delta$) of 10 MHz and (d) its FFT. It survives longer than 20 $\mu$s, which results in a sharp peak in the FFT. (e) The zero quantum coherence with detuning of 5 MHz, and its FFT (f). The two different detuning lead to the same frequency. (g) The mixture of single and zero quantum coherence excited by $\pi/2$ pulse rather than $\pi$ pulse. The FFT (h) shows multiple peaks. The two triples are from single quantum coherences and the single peak near 6 MHz corresponds to the zero quantum coherence ($\Delta$). The 6 MHz is the splitting between the two triplets. }
\label{FIGS3}
\end{figure}

\begin{itemize}
\item Step 1: Initial state : $|\psi(0)\rangle = |1\rangle = \cos{\lambda} |B\rangle + \sin{\lambda} |D\rangle$
\item Step 2: 1$^{\mathrm{st}}$ $\pi$ pulse : $ U_{\pi} |\psi(0)\rangle = \cos{\lambda}|2\rangle + \sin{\lambda} |D\rangle$
\item Step 3:. $e^{i\mathcal{H}_0^R \tau} U_{\pi} |\psi(0)\rangle = \cos{\lambda}e^{-i\delta\tau}|2\rangle + \sin{\lambda}\left(\sin{\lambda}|1\rangle - e^{i\Delta \tau}\cos{\lambda}|2\rangle\right)$
\item Step 4: 2$^{\mathrm{nd}}$ $\pi$ pulse : $|\psi(\tau)\rangle = U_{\pi} e^{-i\mathcal{H}_0^R \tau} U_{\pi} |\psi(0)\rangle
 = \cos{\lambda}e^{-i\delta \tau}|B\rangle + \sin^2{\lambda}\cos{\lambda}(1-e^{i\Delta\tau})|2\rangle + \sin{\lambda}(\sin^2{\lambda} + \cos^2{\lambda} e^{i\Delta\tau})|D\rangle$
\item Sep 5: $p_{2|1}(\tau) = |\langle \psi(\tau)|2\rangle|^2 = 2 \sin^4{\lambda}\cos^2{\lambda}(1 - \cos{(\Delta\tau)})$
\end{itemize}
From the same calculation steps but with the different initial state $|3\rangle$, the formula $p_{2|3} (t)=2 \sin^2{\lambda} \cos^4⁡{\lambda} (1-\cos{(\Delta\tau)})$ can be obtained as well. Finally the $p(\tau)$ is derived as,
\begin{equation}
p(\tau) = \frac{1}{2} + \frac{1}{2}\left(\frac{\Omega_+^4+\Omega_-^4}{(\Omega_+^2+\Omega_-^2)^2}+
2\left(\frac{\Omega_+\Omega_-}{\Omega_+^2+\Omega_-^2}\right)^2\cos{(\Delta\tau)}\right)
\label{EqS9}
\end{equation}

Above equation is close to the Eq.(\ref{EqS8}) except that the oscillation amplitude is half of that in Eq.(\ref{EqS8}). This is because the excitation from the initial states,  $|1\rangle$ or $|3\rangle$, to the state $|2\rangle$ does not happen all the time. The expected success of the excitation, expressed as $p_1(0) |\langle1|B\rangle|^2$ $+ p_3(0) |\langle3|B\rangle|^2$, is $\frac{1}{2}$. This effectively reduces the oscillation amplitude. In the case of $N$ nuclear spins, we can rewrite the Eq.~\ref{EqS9} as $p(\tau)=\left(1-\frac{1}{2^N}\right)+\frac{1}{2^N}  p^{[N]}(\tau)$.  But, the renormalized zero-quantum Ramsey fringe of the system with $N$ nuclear spins, $p^{[N]} (\tau)$ , is the same formula as the Eq.(\ref{EqS8}). This implies that except the less success rate of the transition induced by the mw excitation, reversing the energy level structure upside down makes essentially no differences.

\subsection{Basic properties of zero-quantum coherence}
\subsubsection{Coherence time}
As described in the main text, the zero quantum coherence in the system is mainly on the $^{13}$C nuclear spin. So, it’s expected to have a longer coherence time. Figure \ref{FIGS3} shows a comparison between the single quantum coherence and the zero quantum Ramsey fringes. Figure \ref{FIGS3} (a) and (b) are the single quantum Ramsey fringe detuned by 10 MHz and its Fourier transformation, respectively. We can see that the $T_2^{\ast} (SQ)$ is just about 1 $\mu$s. As shown in Fig.~\ref{FIGS3} (c), the oscillation of the zero quantum Ramsey lasts over 20 $\mu$s. The linewidth of the FFT in Fig.~\ref{FIGS3}(d) is less than 0.1 MHz. This long coherence time of the zero quantum Ramsey fringe leads to the enhanced precision for the estimations of non-secular hyperfine terms.

\subsubsection{Detune independency}
According to the Eq.(\ref{EqS8}), the zero quantum Ramsey fringe is independent of the frequency detuning ($\delta$) of the mw excitation. The results in Fig.~\ref{FIGS3} (c) and (e) confirm this. The curves in the (c) and (e) were detuned by 10 MHz and 5 MHz, respectively. The FFTs are the same as shown in (d) and (f). This detune independency of the measured oscillation is an evidence that the coherence measured is the zero quantum coherence, not single quantum. If it’s a single quantum coherence, it should be dependent on the frequency detuning.

\subsubsection{Excitation pulse duration}
The optimal excitation and readout pulse for the zero quantum Ramsey fringe is the $\pi$ pulse. If one use a different pulse duration, such as $\pi/2$ pulse, the mixture of single quantum and zero quantum coherences appears. One example is shown in Fig.~\ref{FIGS3} (g), in which the $\pi/3$ pulse was used in the sequence. The FFT spectrum shows a zero quantum coherence near 6 MHz in addition to two triplets from 14 to 26 MHz. Single triplet stems from the hyperfine splitting due to the $^{14}$N nuclear spin. The two triplets are split by the amount of the zero quantum frequency, i.e. 6 MHz, which indicates the simultaneous transition of $\Omega_+$ and $\Omega_-$. The disadvantage of the excitation of such multiple coherences is that, a higher number of average is required for the same signal to noise ratio as that of the zero quantum coherence alone. For instance, the curve in (c) was obtained in 30 minutes, but the curve in (g) in 5 hours.

\section{Perturbation calculation of the frequency of zero-quantum Ramsey fringe}
In the system of single NV center with single $^{13}$C nuclear spin, the energy splitting in the ground state $m_S=0$ can be measured from the frequency of zero-quantum Ramsey fringe. The perturbation calculation reveals the contributions of hyperfine interaction components in the system Hamiltonian $\mathcal{H}_0$ to the splitting of the ground state. This section will be devoted to the description of the calculation procedure and the related physical explanations.

\subsection{The simple case with $A_{xx}$, $A_{yy}$, $A_{zz}$}
Before dealing with the actual hyperfine interaction, we start with a simpler form only with $A_{xx}$, $A_{yy}$, and $A_{zz}$. The system Hamiltonian, thus, can be written as
\begin{eqnarray}
\mathcal{H}_0 = D S_z^2 &+& \gamma_e B ( \sin{\theta} \cos{\phi} S_x + \sin{\theta} \sin{\phi} S_y + \cos{\theta} S_z ) + A_{xx} S_x I_x + A_{yy} S_y I_y + A_{zz} S_z I_z \nonumber \\
&+& \gamma_n B ( \sin{\theta} \cos{\phi} I_x + \sin{\theta} \sin{\phi} I_y + \cos{\theta} I_z ),
\end{eqnarray}
in which $\gamma_e$ and $\gamma_n$ are gyromagnetic ratios of electron and nuclear spins, and $D$ zero-field splitting of electron spin of the NV center. The two angles $\theta$ and $\phi$ define the orientation of external magnetic field of which strength is $B$. To get the matrix representation of the Hamiltonian, (6 by 6) one may consider the six states $|m_S, m_I\rangle$ ($m_S=\pm1$, $0$, and $m_I=\pm 1/2$) as the basis. This, however, is not a good choice because for the sub-matrix (2 by 2) of the $\mathcal{H}_0$, represented in the basis of the two states $|0, \pm 1/2\rangle$, the off diagonal component can be as large as diagonal ones, shown as below,

\begin{eqnarray}
\mathrm{Off-diagonal} &:& \langle0, \pm\frac{1}{2}| \mathcal{H}_0|0, \mp\frac{1}{2}\rangle = \frac{\gamma_n B \sin{\theta}}{2} ( \cos{\phi} \pm i \sin{\phi}),\nonumber \\
\mathrm{Diagonal} &:& \langle0, \pm\frac{1}{2}| \mathcal{H}_0|0, \pm\frac{1}{2}\rangle = \frac{\gamma_n B \cos{\theta}}{2}.
\end{eqnarray}

This is because, for nuclear spin states belonging to $m_S=0$, the hyperfine interaction becomes ineffective in first order. The Zeeman energy for nuclear spin, therefore, brings non-negligible off-diagonal components when the external magnetic field is tilted from the $Z$ axis. So, we take a different basis for the matrix representation of the Hamiltonian, which is $|0,\beta_{\pm}\rangle$ rather than $|0,\pm\frac{1}{2}\rangle$. The nuclear spin states $|\beta_{\pm}\rangle$ are eigen states of the nuclear spin Zeeman term, i.e. $|\beta_+\rangle = \cos{\frac{\theta}{2}}|+\frac{1}{2}\rangle + e^{i\phi}\sin{\frac{\theta}{2}}|-\frac{1}{2}\rangle$, $|\beta_-\rangle = \sin{\frac{\theta}{2}}|+\frac{1}{2}\rangle - e^{i\phi}\cos{\frac{\theta}{2}}|-\frac{1}{2}\rangle$. For convenience, we rename the six basis states as below,
\begin{eqnarray}
|1\rangle = |+1, +\frac{1}{2}\rangle&,& |2\rangle = |+1, -\frac{1}{2}\rangle, \nonumber\\
|3\rangle = |0, \beta_+\rangle &,& |4\rangle = |0, \beta_-\rangle, \nonumber\\
|5\rangle = |-1, +\frac{1}{2}\rangle &,& |6\rangle = |-1, -\frac{1}{2}\rangle.
\end{eqnarray}

With this proper basis set, one can straightforwardly obtain the analytic expression of the energy difference between states $|3\rangle$ and $|4\rangle$ from the perturbation calculation of $\mathcal{E}_{|3\rangle} - \mathcal{E}_{|4\rangle}$  ($\equiv \Delta$), ($\Delta \approx \sum_{i(\neq 3,4)}^6 \frac{|\langle i|\mathcal{H}|3\rangle|^2 -
|\langle i|\mathcal{H}|4\rangle|^2}{\langle i|\mathcal{H}|i\rangle}$ ).  The nuclear Zeeman energy contributions can be ignored eventually, and the below is the final expression without higher order contributions Ο$(\frac{A_{xx}^2}{D^2}, \frac{A_{yy}^2}{D^2})$.
\begin{equation}
\Delta = \frac{2 \gamma_e B \sin{\theta}}{D}\left( A_{xx} \cos^2{\phi}+ A_{yy} \sin^2{\phi} \right)
\end{equation}

\subsection{Effect of $aS_z I_x$}
The hyperfine interaction terms in the Hamiltonian considered above are not complete since the two terms $a(S_z I_x+S_x I_z)$ are missing. This section will deal with the effect of the hyperfine interaction $aS_z I_x$ on the eigen states of the Hamiltonian.

For the states associated with $m_S=\pm1$, i.e., $|1\rangle$, $|2\rangle$, $|5\rangle$, and $|6\rangle$, the effect of the hyperfine term $A_{zz} S_z I_z+a S_z I_x$ is dominant in the first order; the other terms make contributions in second order. The four basis states are the eigen states of $A_{zz} S_z I_z$. However, the $a S_z I_x$ modifies the quantization axis of nuclear spin. We can introduce new four basis states such as $|1'\rangle = |+1, \alpha_+\rangle$, $|2'\rangle = |+1, \alpha_- \rangle$, $|5'\rangle = |-1,\alpha_+ \rangle$, and $|6'\rangle = |-1, \alpha_- \rangle$. Here the nuclear spin states
 $|\alpha_{\pm}\rangle$ are eigen states of $A_{zz} I_z+a I_x$. Rewriting the $A_{zz} I_z+a I_x$ as $A'_{zz} (\cos{\theta'} I_z + \sin{\theta'} I_x)$, ($A'_{zz}=\sqrt{A_{zz}^2+a^2}$, $\tan{\theta'} = a/A_{zz}$) gives us the expression of the states $|\alpha_{\pm}\rangle$, $|\alpha_+\rangle = \cos{\frac{\theta'}{2}}|+\frac{1}{2}\rangle + \sin{\frac{\theta'}{2}}|-\frac{1}{2}\rangle$ and $|\alpha_-\rangle = \sin{\frac{\theta'}{2}}|+\frac{1}{2}\rangle - \cos{\frac{\theta'}{2}}|-\frac{1}{2}\rangle$.
In the new basis of the four states, $|1'\rangle$, $|2'\rangle$, $|4'\rangle$, and $|5'\rangle$ the hyperfine interaction $A_{zz} S_z I_z + aS_z I_x$ effectively becomes $A'_{zz} S_z I_z$.

\subsection{Effect of $a S_x I_z$}
Here, we will consider the effect of the hyperfine term $a S_x I_z$, together with $A_{xx} S_x I_x$, on the eigen states of the Hamiltonian. To simplify the arguments, we assume that the external magnetic field is aligned to the NV direction, i.e., $\theta=\phi=0$. In this case, the off-diagonal elements associated with the states $|1\rangle$, and $|2\rangle$, $|3\rangle$, and $|4\rangle$ can be expressed in a matrix form as
\begin{equation}
\left( \begin{array}{cc}
\langle1 | \mathcal{H} | 3\rangle & \langle1 | \mathcal{H} | 4\rangle \\
\langle2 | \mathcal{H} | 3\rangle & \langle2 | \mathcal{H} | 4\rangle  \\
\end{array}\right)
 = \left(\begin{array}{cc}
 a & A_{xx} \\
 A_{xx} & -a \\
 \end{array} \right)
 \end{equation}

Similar to the Sec. III. B, one can modify the two state $|3\rangle$ and $|4\rangle$ to be $|3''\rangle = |0, \eta_+\rangle$ and $|4''\rangle = |0, \eta_-\rangle$, respectively, where $|\eta_+\rangle = \cos{\frac{\theta''}{2}}|+\frac{1}{2}\rangle + \sin{\frac{\theta''}{2}}|-\frac{1}{2}\rangle$ and $|\eta_-\rangle = \sin{\frac{\theta''}{2}}|+\frac{1}{2}\rangle - \cos{\frac{\theta''}{2}}|-\frac{1}{2}\rangle$. If the $\theta''$ satisfies the condition $\tan{\frac{\theta''}{2}} = -\frac{a}{A_{xx}}$, the matrix form above becomes simpler as
\begin{equation}
\left( \begin{array}{cc}
\langle1 | \mathcal{H} | 3''\rangle & \langle1 | \mathcal{H} | 4''\rangle \\
\langle2 | \mathcal{H} | 3''\rangle & \langle2 | \mathcal{H} | 4''\rangle  \\
\end{array}\right)
 = \left(\begin{array}{cc}
 0 & \sqrt{A_{xx}^2 + a^2} \\
 \sqrt{A_{xx}^2 + a^2} & 0 \\
 \end{array} \right)
 \end{equation}
The same argument holds to the case of the other off-diagonal elements associated with the state $|3''\rangle$, $|4''\rangle$, $|5\rangle$, and $|6\rangle$. This implies that the presence of $a S_x I_z$ additional to $A_{xx} S_x I_x$ rotates two of the basis states $|3\rangle$ and $|4\rangle$ by the amount of the $-\theta''$ along $y$ direction, and with the new basis the hyperfine interaction $A_{xx} S_x I_x+a S_x I_z$ can be rewritten simply as $A''_{xx} S_x I_x$. ($A''_{xx} = \sqrt{A_{xx}^2+a^2}$).

\begin{figure}[t]
\includegraphics[width=12cm]{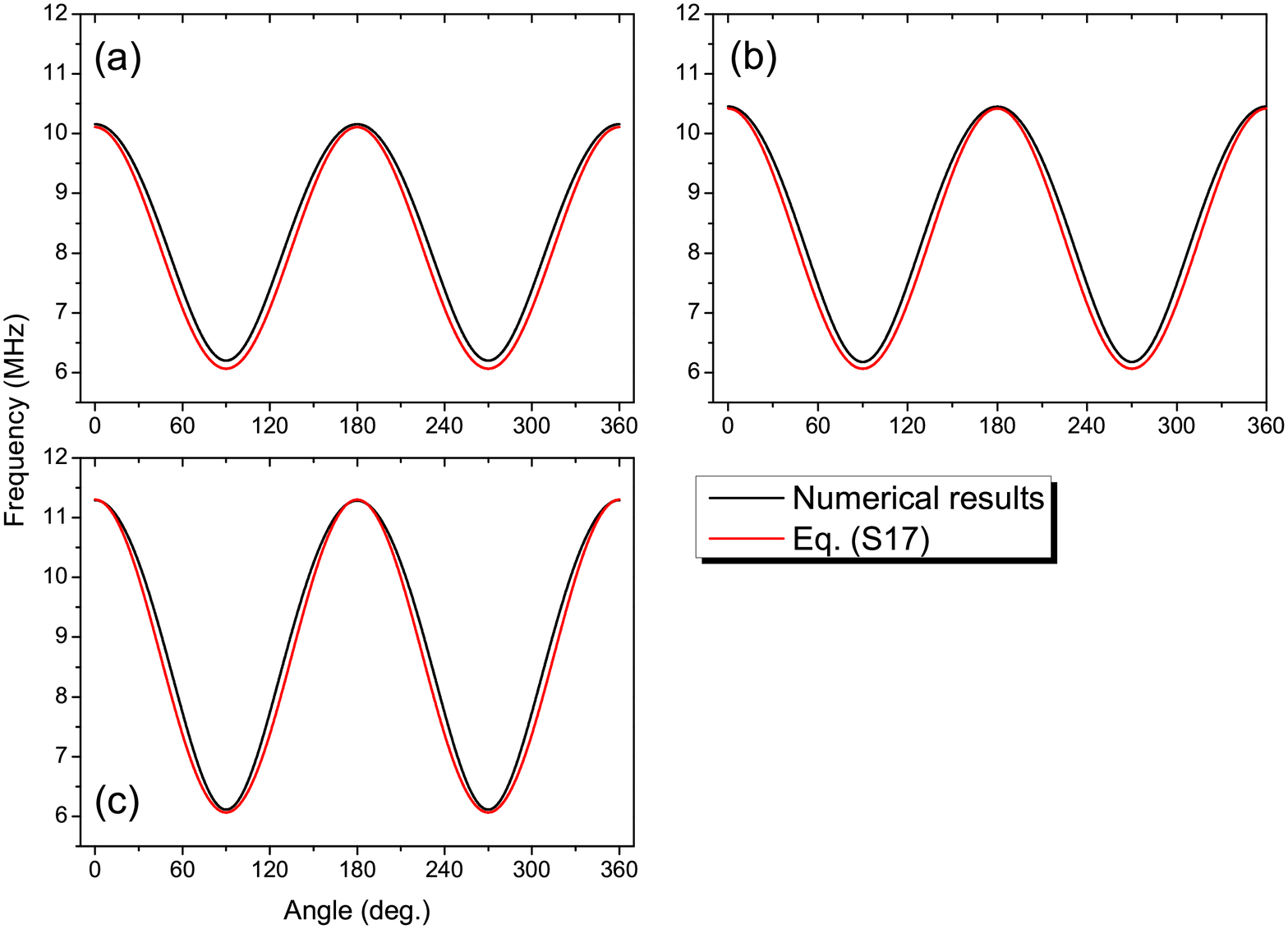}
\caption{(color online)\textbf{Comparison of numerical calculation and analytic expression} For simplicity, the $A_{xx}$ was fixed to be 200 MHz. Then, the a was varied from 0 (a), 50 (b), to 100 MHz (c). The analytic expression is in a good agreement with the numerical results.  }
\label{FIGS4}
\end{figure}

\subsection{The comprehensive case including all the terms}
As described in the main text, according to the symmetry of a single NV center with a nuclear spin, the Hamiltonian is expected to be
\begin{eqnarray}
\mathcal{H}_0 = D S_z^2 &+& \gamma_e B ( \sin{\theta} \cos{\phi} S_x + \sin{\theta} \sin{\phi} S_y + \cos{\theta} S_z ) + A_{xx} S_x I_x + A_{yy} S_y I_y + A_{zz} S_z I_z +a (S_z I_x + S_x I_z)\nonumber \\
&+& \gamma_n B ( \sin{\theta} \cos{\phi} I_x + \sin{\theta} \sin{\phi} I_y + \cos{\theta} I_z ),
\end{eqnarray}

So far in the prior sections, III.A, III.B, III.C, we have handled the influences of the hyperfine interaction terms on the eigen states of the Hamiltonian separately. The notable point is, the quantum states $|m_S\rangle$ properly describe the electron states as long as the external magnetic field is relatively weak. The nuclear spin state, however, are strongly modified due to the presence of the $a (S_z I_x+S_x I_z)$ and due to the rotation of external magnetic field.
Hereby, we conjecture the analytic expression of the $\Delta$ as below,
\begin{equation}
\Delta = \frac{2 \gamma_e B \sin{\theta}}{D}\left( \sqrt{A_{xx}^2+a^2} \cos^2{\phi}+ A_{yy} \sin^2{\phi} \right).
\label{EqS17}
\end{equation}
In Fig.~\ref{FIGS4}, we compared the Eq.(\ref{EqS17}) and the numerical calculation results. For the sake of convenience, the simple values were taken as $A_{xx}$=200 MHz, $A_{yy}$=120 MHz, and $A_{zz}$=130 MHz, for the numerical calculation. The results are for the three values of the $a$, (a) for 0, (b) for 50 MHz, and (c) for 100 MHz. We can see that the expression in Eq.(\ref{EqS17}) is in a good agreement with the numerical results.

\begin{figure}[t]
\includegraphics[width=10cm]{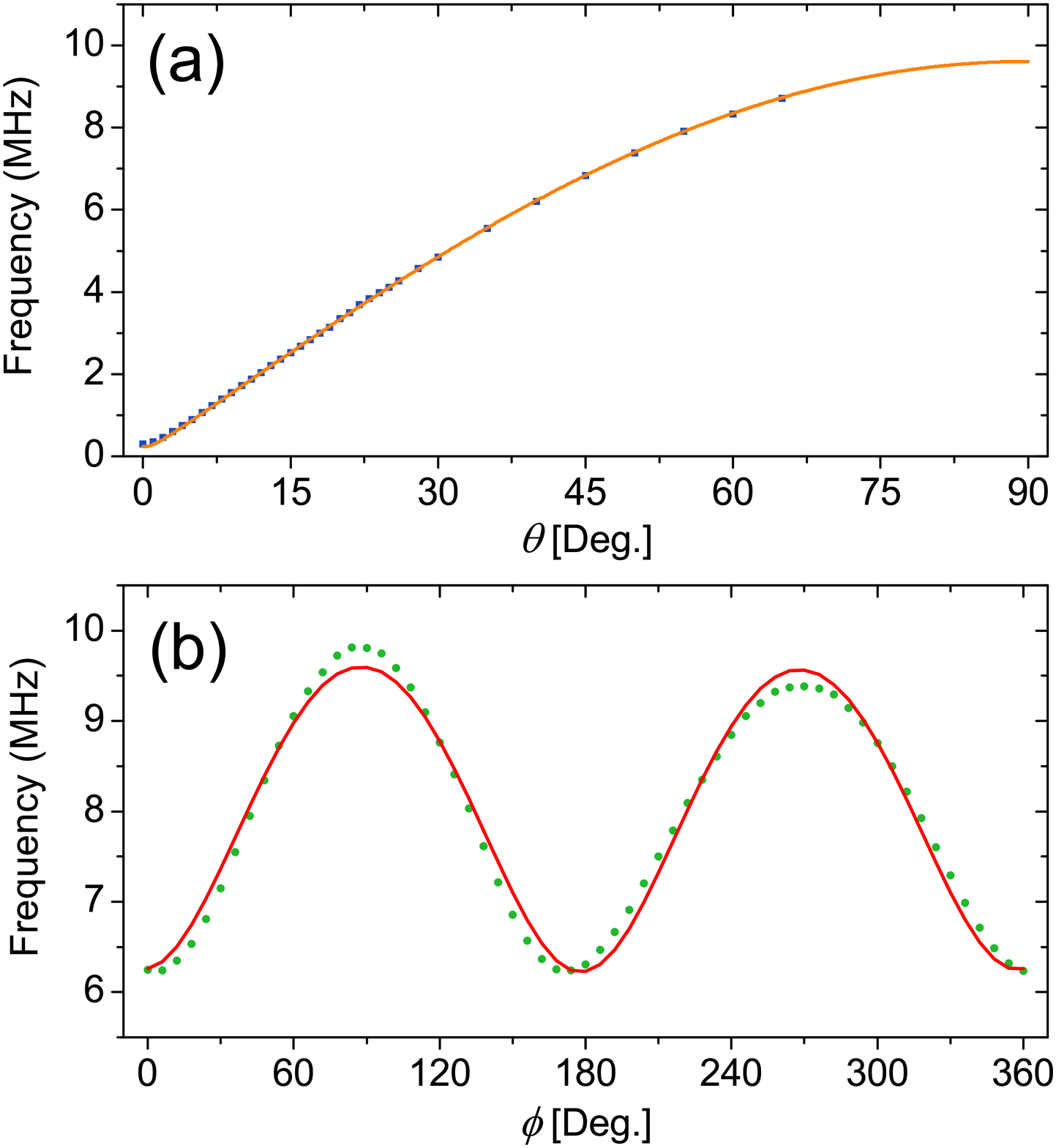}
\caption{(color online)\textbf{Angular variation of the frequency of zero-quantum Ramsey fringe} The frequencies of zero quantum Ramsey fringe were measured as a function of the two angles, $\theta$ and $\phi$. (a) The $\phi$ angle variation from $0$ to $65^{\circ}$ with a fixed $\phi = 90^{\circ}$. (b) The $\phi$ angle variation from 0 to 360$^{\circ}$.}
\label{FIGS5}
\end{figure}

\section{Angular variations of zero-quantum Ramsey fringe}
Figure \ref{FIGS5} confirms that the frequency of zero quantum Ramsey fringe varies as a function of the two angles, $\theta$, $\phi$ according to the Eq.(\ref{EqS17}). The dots are experimental data, and solid lines are numerical fits with the estimated values in the Table I. The result of the $\theta$ rotation at $\phi=90^{\circ}$ is shown in Fig.~\ref{FIGS5} (a), and the $\phi$ at $\theta=40^{\circ}$ in Fig.~\ref{FIGS5} (b). They are clearly consistent with the Eq.(\ref{EqS17}), showing $\sin{\theta}$ and $\cos^2{\phi}$ like curves. Magnetic field strength was fixed to 40.3 Gauss. The deviation between data and the fitting in the $\phi$ rotation will be discussed below.

\section{Rabi oscillation in a $\Lambda$ level structure}
\begin{figure}[t]
\includegraphics[width=12cm]{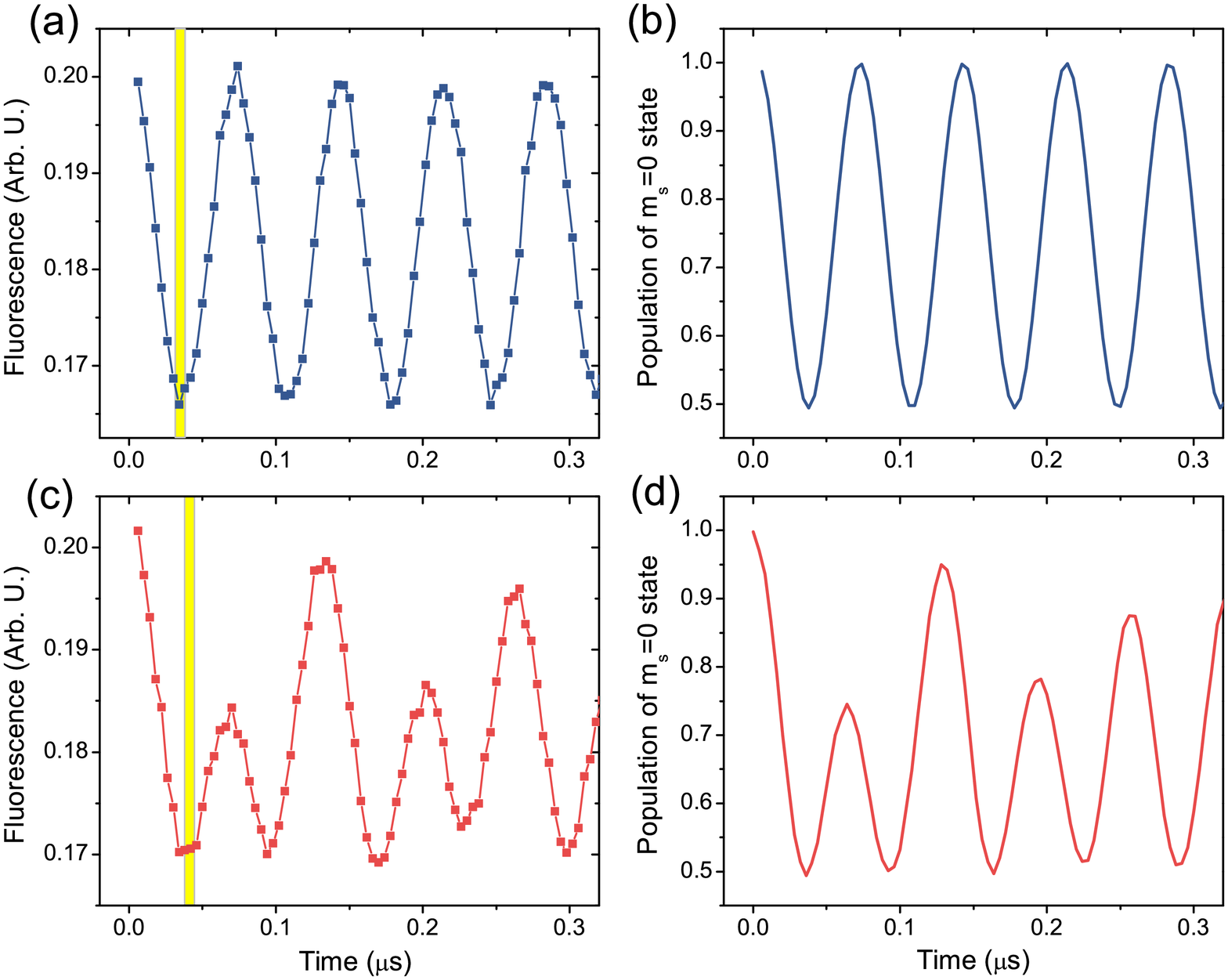}
\caption{(color online)\textbf{The measured Rabi oscillations and simulations} The Rabi oscillations measured under the single transition axis (a) and a tilted orientation (c). The curve in (a) shows a conventional behavior while that in (c) has a modulation. The simulation results well reproduces the experimental results (b) and (d). The yellow bars indicate the durations of $\pi$ pulses. }
\label{FIGS6}
\end{figure}

When the mw field induces the two transition simultaneously, the Rabi oscillation in a $\Lambda$ structure will be unlike to the case of the single transition axis, where only one transition is involved. Figure \ref{FIGS6} compares the two cases. The blue corresponds to the single transition axis, while the red to the tilted orientation, precisely $\theta=40^{\circ}$ and $\phi=90^{\circ}$ in the NV frame (Fig. 1(a) in the main text). The Rabi oscillation under the single transition axis shows a conventional sinusoidal behavior in Fig.~\ref{FIGS6} (a). But, for the tilted case, it shows a modulation as in Fig.~\ref{FIGS6} (c). From the curve in Fig.~\ref{FIGS6} (a), we can obtain the strength of the mw field from the Rabi frequency, which is 14.3 MHz. With this strength, we performed the simulation based on the obtained hyperfine parameters in the main text (Table I) and Fig.~\ref{FIGS6} (b) and (d) are the results. The simulation reproduces the obtained results in good agreements.

The durations of the $\pi$ pulses used for the zero quantum Ramsey sequence are marked as yellow in Fig.~\ref{FIGS6} (a) and (c). We can notice that they are nearly identical. In the experiments, the times of the first minima of Rabi oscillations were taken as the durations of the $\pi$ pulses.

\section{Dependency of single quantum spectrum on hyperfine parameters}
\begin{figure}[t]
\includegraphics[width=12cm]{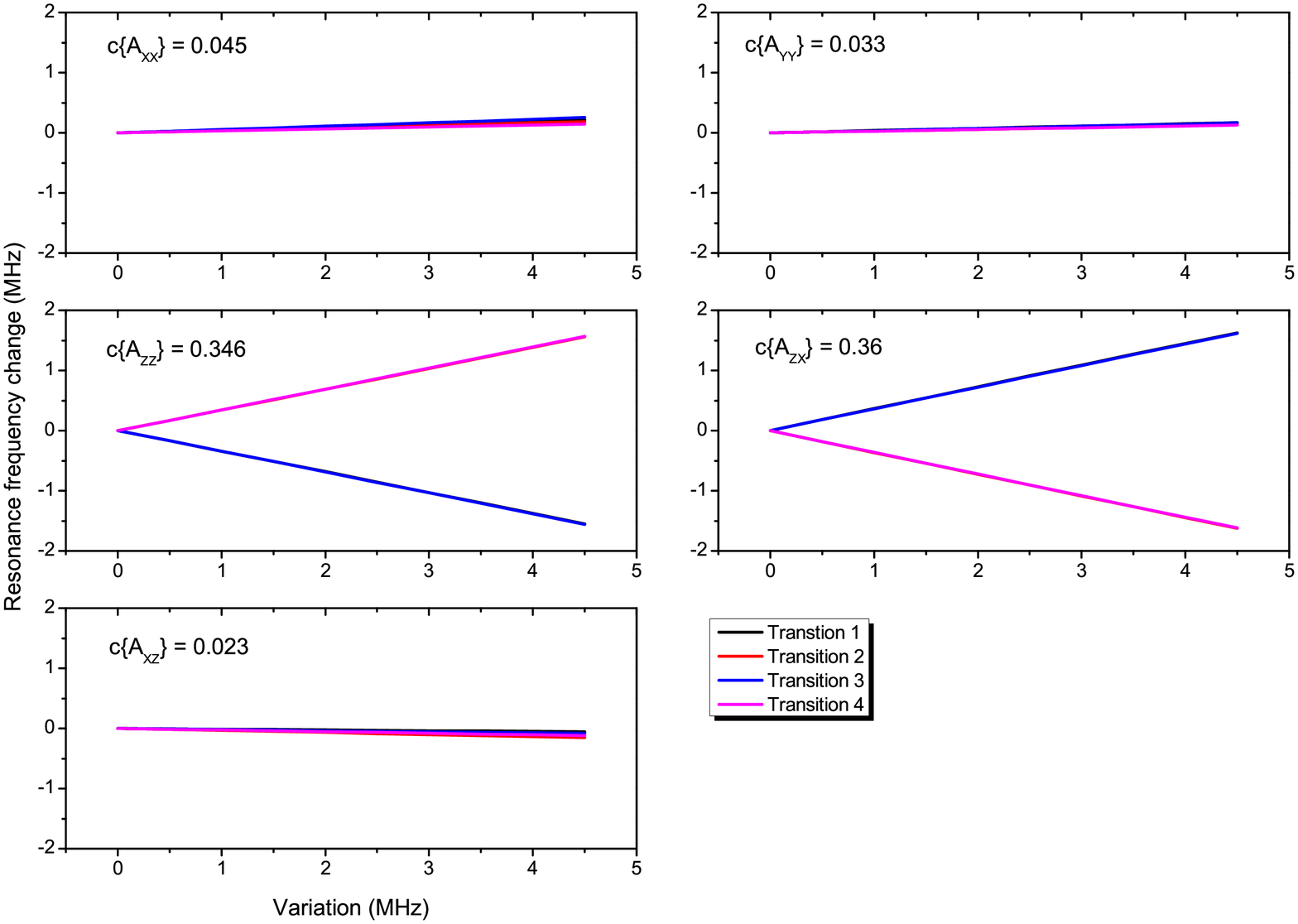}
\caption{(color online)\textbf{The dependencies of transition frequencies on hyperfine parameters} At the condition of the single transition axis, the variation of the four allowed transition frequencies were numerically calculated by varying hyperfine parameters, $A_{xx}$, $A_{yy}$, $A_{zz}$, $A_{zx}$, $A_{xz}$. Only single parameter was varied while keeping the others fixed as the values in the Table I of the main text. The values $c{A}$ are the average slope of the four transitions. The solid lines were obtained from numerical calculation with the parameters in the Table I.}
\label{FIGS7}
\end{figure}

The transition frequencies in the single quantum spectrum as shown in Fig. 3(a) are not sensitive to all the parameters in the hyperfine interaction. As explained in the main text, the $c\{A\}=\frac{\partial \omega}{\partial A}$ reveals the degree of the dependence, which is related to the precision in the estimation of the parameter $A$. If one can measure the transition frequency $\omega$ with a precision of $\delta \omega$, the parameter $A$ can be estimated with a precision of $\delta A = \delta \omega / c$.
Figure \ref{FIGS7} shows the variations of the four allowed transition frequencies under the single transition axis for the five parameters. Numerical calculation were performed and the initial values adopted in the calculation were the same as in the Table I in the main text. For each parameter, the average slope of the four transitions is displayed. One can notice that the $c\{A_{zz}\}$and the $c\{A_{zx}\}$ are an order of magnitude large than the others.

\section{Single transition axis}
\begin{figure}[t]
\includegraphics[width=14cm]{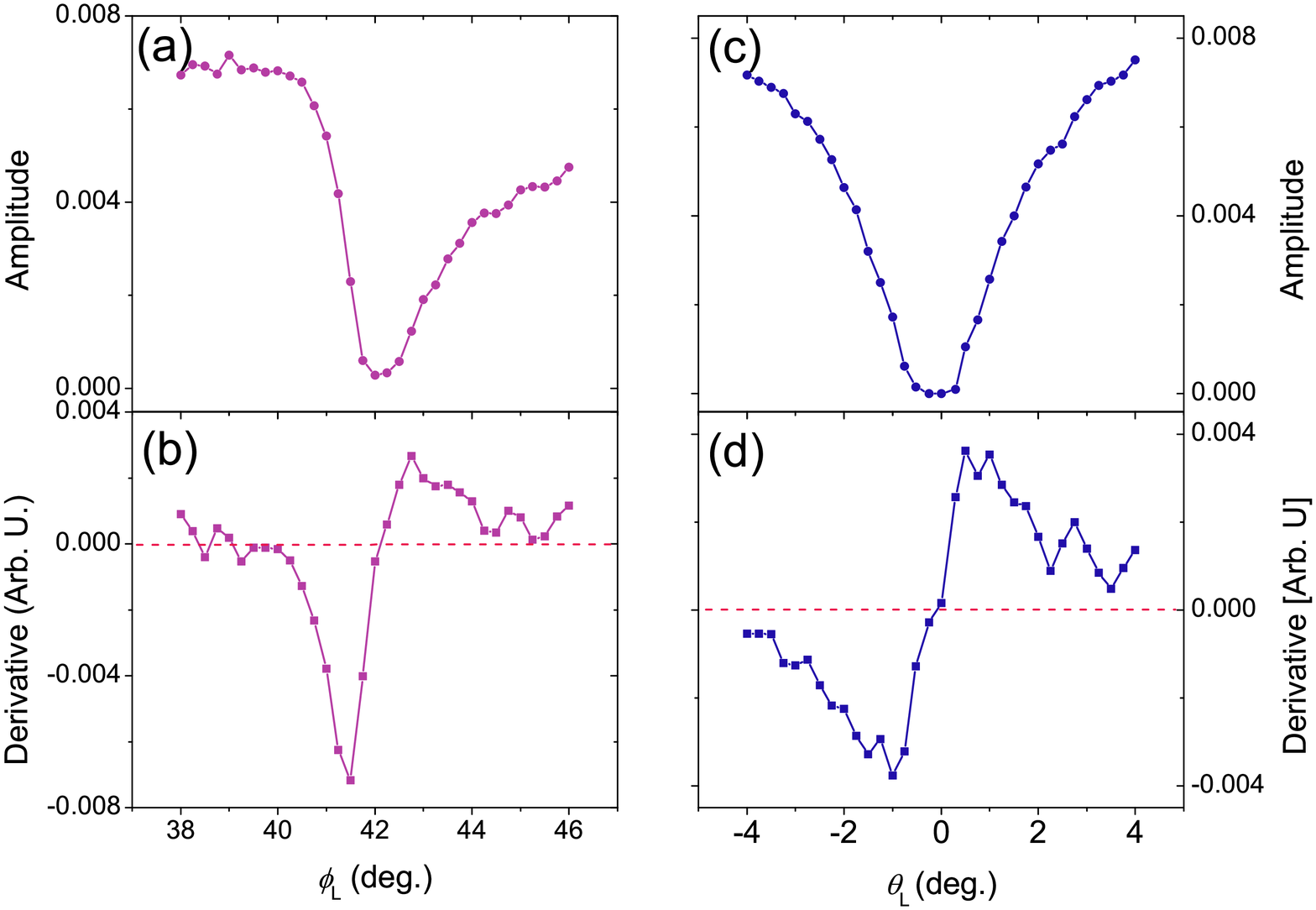}
\caption{(color online)\textbf{The orientation of the single transition axis in the Lab. frame} In (a) and (c), the intensities of the measured oscillations in the Fig. 2 (c) and (d) are plotted. The direction of the single transition axis can be assigned to the orientation having the minimal oscillation amplitude. (b) and (d) are the derivatives of the (a) and (c). The dashed red lines indicate zero, and the crossing points near 42$^{\circ}$ in (b) and near 0$^{\circ}$ in (d) will give the orientation of the single transition axis.}
\label{FIGS8}
\end{figure}

The orientation of the single transition axis can be detected from the variation of the strength of the zero quantum Ramsey fringe as a function of the two angles, $\phi_L$ and $\theta_L$ in the Lab.~frame. According the Eq.(\ref{EqS8}), the zero quantum Ramsey should present no oscillation because one of the transition amplitude, either $\Omega_+$  or $\Omega_-$ vanishes under the single transition axis. In Fig.~\ref{FIGS8} (a) and (c), the amplitudes of the oscillations shown in the Fig.3(b) are plotted. We tried to find the angle of the minimum oscillation amplitude by performing derivative of the data and picking the point closest to the zero lines (red dash lines).

\section{Imperfections in the rotation of magentic field}
\begin{figure}[h]
\includegraphics[width=\textwidth]{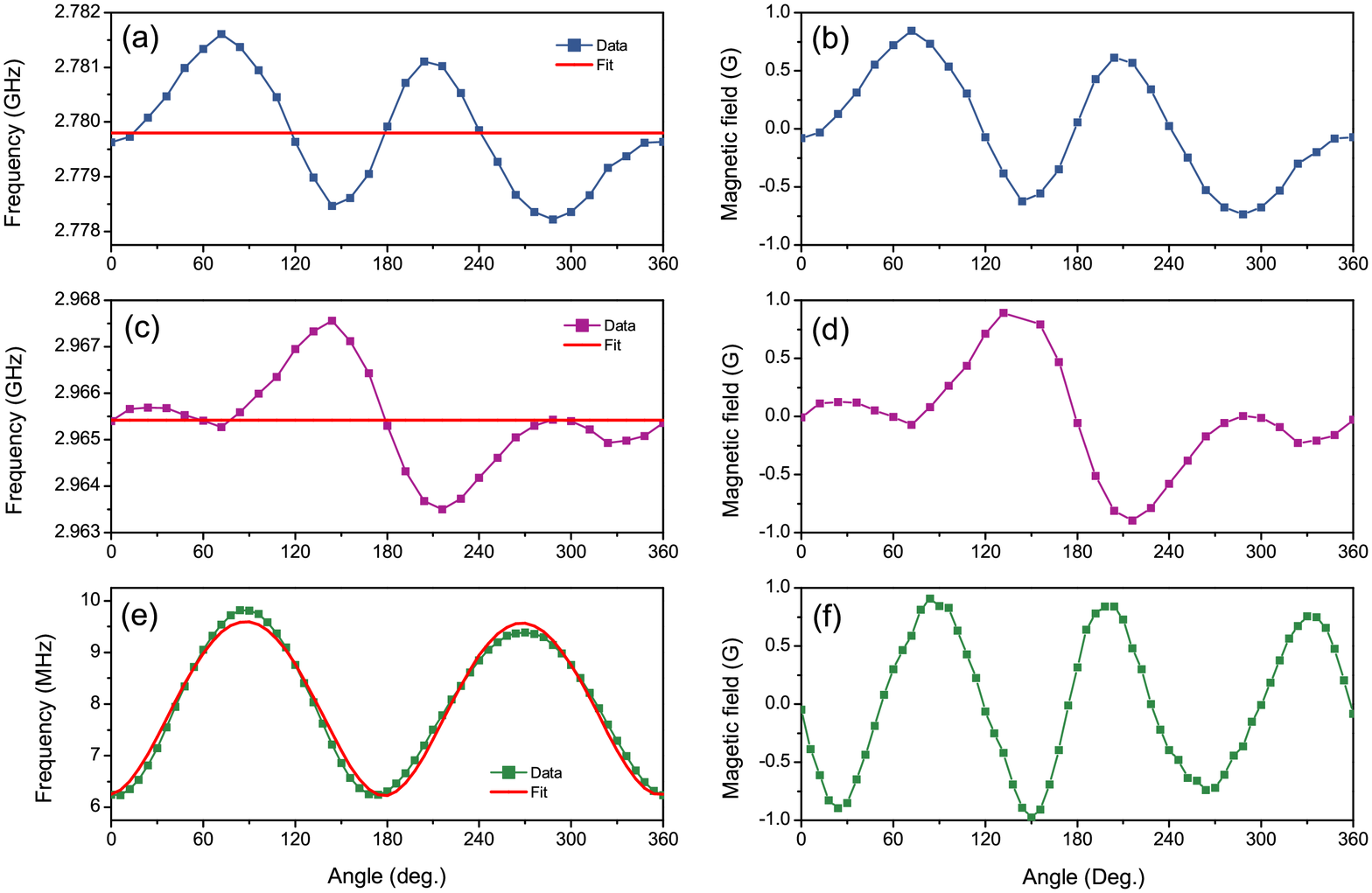}
\caption{(color online) \textbf{Imperfection in the $\phi$ rotation} The frequencies of the zero quantum coherence are plotted as a function of the $\phi$ angle in (e) and the deviation from the numerical fit is shown in (f). To test the variation of the magnetic field strength during the $\phi$ rotation, the same rotation was performed in a nearby NV center having the same crystallographic orientation. The curves in (a) and (c) correspond to the two transitions and the deviations are in (b) and (d).}
\label{FIGS9}
\end{figure}

Figure \ref{FIGS9} (e) shows the variation of the frequencies of zero-quantum Ramsey fringes with respect to a full $\phi$ angle rotation in the NV frame and the fitted curve resulting from the numerical analysis. The deviation between the data and the fitting is plotted in Fig.~\ref{FIGS9} (f), which is converted from the frequency into the strength of the magnetic field (Gauss) according to the Eq.(\ref{EqS17}). The deviation shows a systematic behavior having a period of around 120 degree. To see if it arises from the imperfection of the rotation of magnetic field, specifically the variation of the magnetic field strength during the rotation, we performed the same rotation on one of the other nearby NV center and measured single quantum Ramsey spectra.

Figure \ref{FIGS9} (a) and (c) display the variation of the transition frequencies of the two transitions. ((a) for $m_S=0 \rightarrow m_S=-1$ and (b) for $m_S=0 \rightarrow m_S=+1$) The red curves are fittings and the deviations in (b) and (d) have similar periods as in (f). The curve in Fig.~\ref{FIGS9} (d) is out of phase with respect to the curve in Fig.~\ref{FIGS9} (b) as expected. Although they are not identical, the amount of magnetic field variation are nearly the same, i.e. $\pm$ 1 G. There can be an imperfection in the $\phi$ angle rotation in our rotation stage. This imperfection produces a non-uniform magnetic field strength during the rotation, which is believed to be the source of the deviation in Fig.~\ref{FIGS9} (f).

\end{document}